%% file: template.tex
\documentclass[preprint,journal]{vgtc}

\input{"packages.tex"}

\input{"commands.tex"}

\onlineid{1437}

\vgtccategory{Research}

\title{PoseForge: Editable Pose Analytics for AI-Assisted Sports Coaching}

\author{%
  \authororcid{Shuvam Swapnil Dash}{0009-0006-6685-5636},
  \authororcid{Arpit Narechania}{0000-0001-6980-3686}
}

\authorfooter{
  \item
  	Shuvam is with Autotake Developers Private Limited and an incoming research intern at The Hong Kong University of Science and Technology.
  	E-mail: shuvamswapnil21@gmail.com
  \item
  	Arpit is with The Hong Kong University of Science and Technology.
    E-mail: arpit@ust.hk
}

\abstract{Athletic coaching increasingly relies on video analysis, yet raw footage lacks tools to quantify motion or simulate valid technique corrections. Drawing on formative interviews with eleven cricket experts (coaches, performance analysts, captains, and players), we introduce PoseForge, a visual analytics system that extracts 3D skeletal poses from single-camera sports videos for interactive movement analysis. In a cricket batting case study, PoseForge computes interpretable kinematic metrics such as feet gap and elbow angle, compares them against scientifically derived norms, and uses an AI coach to suggest targeted adjustments, presented visually and through natural-language feedback (e.g., ``increase feet gap by 10 cm''). Users can directly modify poses via mouse interaction or natural-language instructions, with inverse kinematics maintaining anatomical plausibility and real-time updates of metrics and comparisons. An evaluation with the same eleven cricket experts found PoseForge effective for diagnosing movement issues and exploring corrective alternatives, highlighting its applicability in low-resource, academy, and grassroots coaching settings, while identifying opportunities for enhanced sport-specific metrics and longitudinal tracking. PoseForge is available as open-source software at \url{https://github.com/DataVisards/PoseForge}.}


\teaser{
  \centering
  \includegraphics[alt={Screenshot of the PoseForge interface displaying five functional panels. The layout includes areas for session setup, a batting video viewer, an interactive 3D skeleton editor, a biomechanical metrics dashboard with AI feedback, and the analysis history.}, width=\linewidth]{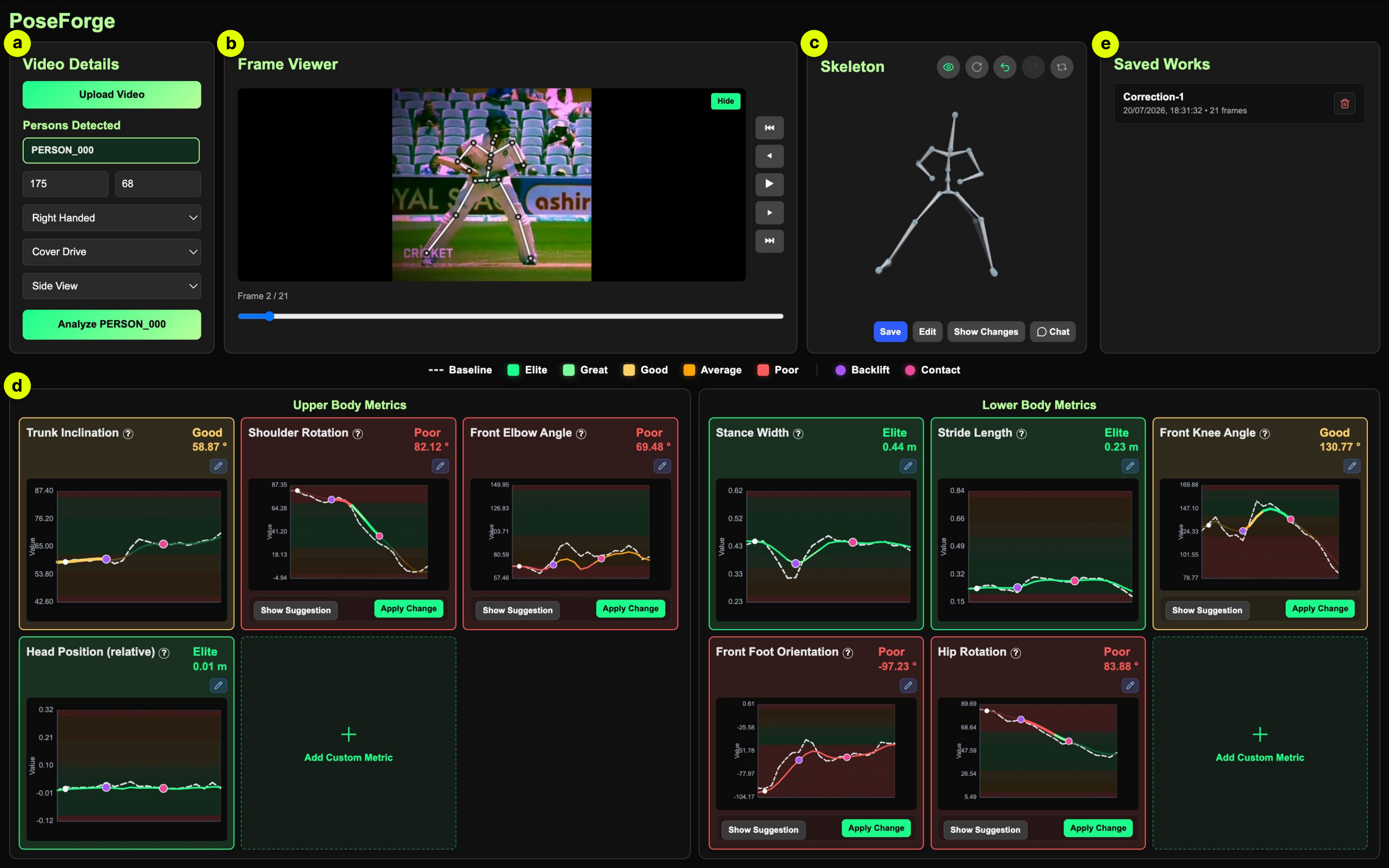}
  \caption{PoseForge interface showing coordinated views for kinematic analysis and editing. Users can (a) configure the session by uploading a video and specifying context, (b) inspect and navigate the input video through the frame viewer, (c) visualise and interactively edit the reconstructed 3D skeleton with real-time feedback, (d) analyse upper- and lower-body metrics with phase-aware grading and AI-generated coaching suggestions, and (e) manage saved corrections and revisit prior analysis sessions.
  \label{fig:teaser}
  }
}

\graphicspath{{figs/}{figures/}{pictures/}{images/}{./}} 

\usepackage{tabu}                      
\usepackage{booktabs}                  
\usepackage{lipsum}                    
\usepackage{mwe}                       
\usepackage{ccicons}                   

\usepackage{mathptmx}                  

\begin{document}



\firstsection{Introduction}

\maketitle 

Athletic excellence manifests in motion, in the precise timing, 
coordination, and sequencing of joints and body segments that 
separate elite performance from developing technique. The 
differences that matter most occur within fractions of a second 
and involve postural deviations too subtle for the naked 
eye~\cite{bartlett2007introduction}. For decades, sports coaches have 
relied on video replay, qualitative observation, and personal 
judgment, methods that remain inherently subjective and 
constrained by single camera angles~\cite{hughes2004overview}. 
Marker-based motion capture and force plates yield high-fidelity 
kinematic data~\cite{windolf2008systematic}, yet remain costly, 
laboratory-bound, and inaccessible to most academies and 
grassroots programs.

In skill-intensive sports such as cricket or 
baseball, these constraints carry both 
practical and clinical consequences. A player may produce an 
excellent outcome, such as a boundary or home run, even with 
suboptimal posture or joint alignment. Such compensatory 
techniques often accumulate stress injuries over 
time~\cite{elliot2000back, orchard2010injuries}. Hence, 
evaluating performance quality should not depend solely on 
outcomes but also on the \textit{biomechanical soundness} of 
the underlying motion~\cite{mcginnis2013biomechanics}. 
Achieving that balance demands tools that reveal whether 
technique was both efficient and safe. Hence, we set a high-level goal \textit{``to help sports coaches complement performance metrics with biomechanical indicators of technique quality and injury risk.''}

To operationalize this goal, we focus on \textbf{cricket}, where a player's pose—captured through kinematic indicators such as knee flexion and elbow orientation—directly influences balance, timing, and shot control~\cite{stuelcken2005offside}. 
While elite settings may leverage motion capture systems or wearable sensors to obtain such measurements, they remain largely inaccessible in grassroots coaching. As a result, coaches routinely assess these cues from live play or video, but the lack of reliable quantification makes such assessments subjective and difficult to reproduce~\cite{lees2002technique, franks1997need}. Moreover, the relevance of these kinematic cues is inherently phase-dependent, with different metrics becoming critical at different stages of a movement. 
We therefore ask: \emph{``How can we generate 3D kinematic reconstructions from single-view coaching videos and derive phase-aware metrics to objectively quantify athletic technique?''}

Building on this, even with access to kinematic data, translating analysis into actionable corrections remains non-trivial. Verbal instructions are often spatially ambiguous~\cite{magill2011motor}, and static annotations fail to capture the continuity of motion. Moreover, suggested corrections must remain biomechanically valid, respecting human joint limits to avoid unrealistic or unsafe movements~\cite{aristidou2011fabrik}. This leads to a second challenge: enabling intuitive and physically plausible exploration of technique adjustments. We thus ask: \textit{``How can we enable interactive, anatomically plausible exploration of corrective postures for grassroots coaching?''}

To address these questions, we first conducted a formative  interview study with eleven cricket experts --- including coaches, performance analysts, and tech-savvy team captains and players --- spanning from grassroots to national levels to understand their analysis routines, communication styles, and existing tool use~\cite{lazar2017research}. 
Based on these findings, we built \textbf{PoseForge}, a visual analytics tool that reconstructs 3D skeletal poses from single-view videos. Using a markerless estimation pipeline~\cite{ferguson2025mhr}, it performs a 3D kinematic reconstruction of the pose before computing interpretable metrics such as feet gap, elbow angle, and body lean, which are normalized and graded against empirically defined thresholds for consistent evaluation.
Coaches can directly adjust poses through mouse interactions or natural language instructions~\cite{brown2020language}, with all edits resolved by an inverse-kinematics constraint engine~\cite{aristidou2011fabrik} that enforces anatomical plausibility and updates metrics in real time. 
An AI assistant offers phase-aware coaching suggestions, while custom metric editors enable sport-specific definitions.

We also conducted a summative evaluation by interviewing the same eleven experts as in the formative interview study. 
Experts responded enthusiastically to PoseForge's interactive and phase-aware features, describing the system as a potential `game changer for grassroots coaching.'
They valued the plausibility of inverse-kinematic edits~\cite{aristidou2011fabrik}, the clarity of quantified displacement overlays, and the accessibility of natural-language corrections~\cite{casiez2012oneeuro}. Additionally, constructive feedback highlighted opportunities for more cricket-specific metric interpretation and longitudinal tracking. Overall, our findings suggest that the barriers separating grassroots coaching from professional-grade motion analysis are getting narrower. A browser-based system powered by single-camera input and coaching-language interaction can make high-fidelity kinematic reasoning accessible, interpretable, and practically useful, amplifying human expertise through physically grounded visualization~\cite{isenberg2011systematic}.
The primary contributions of this work include:

\begin{itemize}[nosep]
    \item Findings from a formative interview study with 
    eleven cricket experts on current coaching, analysis, 
    and communication practices.
    \item \textbf{PoseForge}, a visual analytics tool for cricket experts (e.g., coaches, analysts) to track, analyze, and correct a batter's pose. 
    \item A kinematically constrained \textit{what-if} analysis pipeline enabling hypothetical pose corrections with real-time metric updates.
    \item Feedback from eleven cricket experts about 
    PoseForge's utility and potential to be integrated into real coaching workflows.
\end{itemize}

\section{Related Work}
\subsection{Kinematic and Biomechanical Analysis in Sports}

Quantitative kinematic and biomechanical analysis of athletic movement is a well-established area of study in sports science. Marker-based motion capture and force plates have established typical ranges for joint angles, segment velocities, and ground reaction forces across sports and skill levels~\cite{Davis1991AGA, dufek1990biomechanical}. Augmented feedback has been shown to accelerate skill acquisition and correct persistent technical faults~\cite{sigrist2013augmented,swinnen1996information}. More recently, video-based automated grading systems have demonstrated the feasibility of replacing laboratory infrastructure with computer vision pipelines for routine technique assessment~\cite{worsey2019systematic,lee2025vair}.

Simulation and movement-modification frameworks have also emerged from biomechanics and computer graphics research, including musculoskeletal simulation environments \cite{delp2007opensim} and interactive motion-editing techniques (Section~\ref{sec:motion}). While these approaches support assessment or movement simulation, integrated workflows combining technique grading, corrective editing, and biomechanical evaluation within a unified coaching workflow remain uncommon~\cite{barris2008review}.

\subsection{3D Human Pose Reconstruction and Motion Editing}
\label{sec:motion}
Recent advances in parametric human body models such as
SMPL~\cite{loper2015smpl} and SMPL-X~\cite{pavlakos2019expressive},
together with transformer-based pose estimation methods including
MotionBERT~\cite{zhu2023motionbert} and MixSTE~\cite{zhang2022mixste},
have substantially improved markerless 3D motion reconstruction from
video. Sport-specific datasets such as
SportsPose~\cite{ingwersen2023sportspose} and recent models including
ATLAS~\cite{atlas} and the Momentum Human Rig
(MHR)~\cite{ferguson2025mhr} further improve reconstruction of athletic
motions under challenging conditions such as rapid limb movement and
self-occlusion.
Interactive correction of reconstructed motion has traditionally been
studied through inverse kinematics, space-time constraints~\cite{witkin1988spacetime}, motion warping~\cite{witkin1995motion}, and musculoskeletal simulation frameworks such
as OpenSim~\cite{delp2007opensim}, Visual3D~\cite{visual3d}, and Anybody~\cite{anybody}. Modern editing
interfaces combine direct joint manipulation using 3D transform gizmos~\cite{mine1997moving} with IK solvers such as FABRIK~\cite{aristidou2011fabrik} and temporal smoothing
methods such as the One Euro Filter~\cite{casiez2012oneeuro}. PoseForge integrates these
techniques with sport-specific kinematic constraints, metric computation, and corrective editing within a unified coaching workflow.

\subsection{Visual Analytics for Athletic Performance}

Visual analytics (VA) has a wide range of applications in sports,
from team-level tactical analysis and positional
tracking~\cite{perin2018state, andrienko2017visual} to passing
network visualization~\cite{gudmundsson2017spatio} and individual
athlete performance analysis~\cite{polk2014tennivis,
stein2017bring}. These systems demonstrate the value of interactive visual
interfaces for sports analysis; however, their primary focus is
understanding match dynamics, tactics, and overall performance within
game contexts. PoseForge addresses a different analytical scale by
supporting fine-grained biomechanical assessment and corrective
simulation of individual batting technique. By reconstructing 3D motion
directly from monocular video and exposing phase-aware kinematic
metrics through an interactive coaching interface, it extends
sports visual analytics toward technique-focused coaching and skill
development.

\subsection{AI-Assisted Sports Coaching}
AI-assisted coaching systems have increasingly been explored for
sports skill assessment, combining computer vision, biomechanical
analysis, and machine learning to provide automated technique
evaluation and feedback. Recent systems such as
SportGPT~\cite{tian2025sportsgpt} employ large language models to
generate interpretable coaching advice from estimated body motion,
while sports visual analytics systems incorporate AI-generated
explanations to support practitioner decision
making~\cite{lee2025vair}. However, these approaches typically
provide textual feedback without allowing coaches to interactively
simulate corrective movements or directly manipulate reconstructed
poses. PoseForge closes this loop by combining metric-grounded LLM
feedback with interactive pose editing in a unified coaching
interface.

\begin{figure*}[!ht]
    \centering
    \includegraphics[alt={Block diagram of the PoseForge processing pipeline. The flow progresses from monocular batting video input through pose estimation, metric computation, and interactive pose correction, ending with AI-generated feedback.}, width=0.9\linewidth]{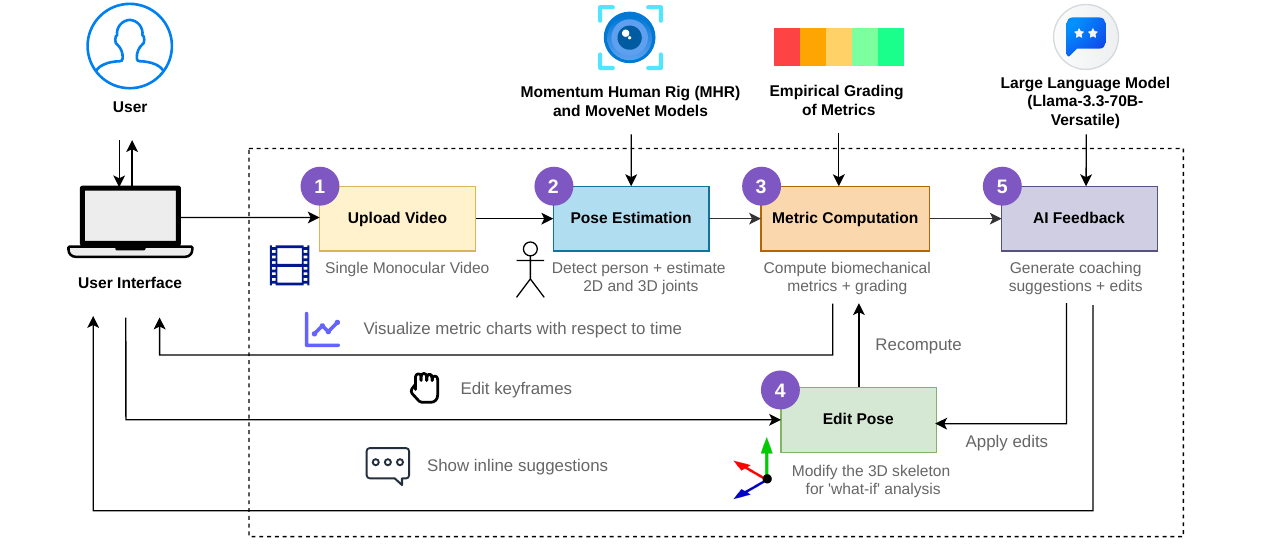}
    \vspace{-16pt}
    \caption{System architecture of PoseForge. A monocular batting video is processed through a pipeline comprising pose estimation, metric computation, pose correction via what-if analysis, and AI-driven feedback generation.}
    \label{fig:architecture}
\end{figure*}

\section{PoseForge - A Case Study with Cricket}

Cricket is a bat-and-ball sport played between two teams, in which one side (the batting team) attempts to score runs, while the other (the fielding team) seeks to limit scoring and dismiss the batters~\cite{woolmer2008art}. The game is played on a 22-yard rectangular pitch, with a wicket positioned at each end; each wicket consists of three vertical wooden stumps topped by two small horizontal bails. During play, a bowler from the fielding team delivers the ball toward the batter, who attempts to strike it and score by running between the wickets or sending the ball to the boundary. The fielding team, comprising the bowler and fielders, works to restrict these scoring opportunities and achieve dismissals through accurate bowling, fielding, and by forcing errors in play.

Batting constitutes the primary offensive skill in cricket and involves executing a diverse range of strokes in response to varying pitch conditions and ball deliveries. These strokes are typically classified based on their direction, timing, and associated footwork, requiring coordinated movement of both the upper and lower body~\cite{cotton2019cricketbiomechanics}. Among these, drive shots—played with a straight or near-straight bat to keep the ball along the ground—are considered fundamental to effective batting technique. In particular, the cover drive, an attacking front-foot stroke directed through the off side of the field, exemplifies the integration of balance, precise foot placement, and controlled bat swing~\cite{stuelcken2005offside}. Understanding the mechanics underlying such strokes provides a useful foundation for analyzing batting technique and developing performance-enhancing tools. For more details on cricket, we refer the reader to~\cite{woolmer2008art}.

\subsection{Formative Interviews with Cricket Experts}
\label{sec:formative_study}
To ground the design of the system in authentic coaching and analytical practice, we conducted semi-structured 
interviews~\cite{lazar2017research} centred around understanding current workflows, 
tools, and unmet needs in technique assessment. Our goal was to first identify gaps and opportunities (formative assessment) that would drive our tool's design and development; after which we would re-interview these experts for their feedback (summative evaluation, Section~\ref{sec:eval-summative}). The entire study was conducted at Autotake Developers Private Limited when the lead author was employed there, and which did not require IRB approval for non-clinical human-subjects research. Informed consent was sought from each participant before the study. Participants were not compensated.

\vspace{0.1cm}\noindent\textbf{Participants.}
We interviewed eleven cricket experts ($P1$--$P11$; 9 males, 2 females) recruited through targeted outreach and snowball sampling~\cite{patton2002qualitative}. These included a performance analyst ($P1$), a captain ($P2$), coaches 
(4: $P3$, $P9$--$P11$) and players (5: $P4$--$P8$). 
Relevant professional experience ranged from 8 to 35 years (median: 10). 
All participants had experience reviewing batting footage and reported familiarity with commercial sports analysis tools as part of their coaching workflows.
They have been involved in cricket-related activities across India, Australia, and Hong Kong--from grassroots and youth programs to national and international competition--across both men’s and women’s formats. Two participants had also played or coached teams at the international level. To preserve anonymity, participants’ names and affiliations are not disclosed.

\vspace{0.1cm}\noindent\textbf{Procedure.}
We conducted individual sessions remotely via Zoom, each lasting approximately 25--30 minutes. Participants first filled a background questionnaire (comprising prompts about their roles, level of athletes they work with, how frequently they analyze player poses, and tools currently used). Then, we asked them the following series of semi-structured questions~\cite{flanagan1954critical}. 

\begin{tcolorbox}[
    breakable,
    colback=gray!20,
    colframe=gray!50,
    boxrule=0.3pt,
    arc=0mm,
    left=1mm,
    right=1mm,
    top=1mm,
    bottom=1mm
]
\footnotesize
\begin{enumerate}[leftmargin=*,nosep]
    \item \textit{``What are the parameters based on which you judge a batter's technique, and how do you know if a parameter is good or bad?''}

    \item \textit{``What methods and tools do you currently use to analyse a batter's technique? What do you like or dislike about them?''}

    \item \textit{``Describe a recent experience in which you noticed a technical error in the strokeplay of a batter. How did you confirm it, determine the correction, and communicate it to the player?''}

    \item \textit{``Did you face any challenges in pinpointing the exact magnitude of change required, or in communicating that to the player?''}
\end{enumerate}
\end{tcolorbox}

\vspace{1mm}

\begin{table}[ht]
\footnotesize
\centering
\caption{Skeletal keypoints used by PoseForge. MHR Index denotes the corresponding joint ID in the MHR pose estimation model. Chain Role indicates how each joint participates in the inverse kinematics pipeline: \emph{Root} anchors the kinematic chain (Pelvis), \emph{chain root} initiates an arm or leg chain (Shoulder, Hip), \emph{hinge} enforces anatomical joint limits (Elbow, Knee), \emph{node} is an intermediate joint (Ankle), and \emph{end-effector} is the manipulated endpoint of a chain (Wrist, Foot). L/R denotes leading/trailing side relative to batting stance.}
\label{tab:keypoints}
\begin{tabular}{lll}
\hline
\textbf{Keypoint} & \textbf{MHR Index} & \textbf{Chain Role} \\
\hline
Pelvis         & 0       & Root \\
Spine1         & 34      & Spinal chain \\
Spine2         & 35      & Spinal chain \\
Spine3         & 36      & Spinal chain / arm branch root \\
Neck           & 109     & Spinal chain \\
Head           & 125     & Spinal chain terminus \\
L/R Shoulder   & 74 / 38 & Arm chain root \\
L/R Elbow      & 75 / 39 & Arm hinge ($8^\circ$--$176^\circ$) \\
L/R Wrist      & 76 / 40 & Arm chain end-effector \\
L/R Hip        & 17 / 1  & Leg chain root \\
L/R Knee       & 18 / 2  & Leg hinge ($6^\circ$--$178^\circ$) \\
L/R Ankle      & 20 / 5  & Leg chain node \\
L/R Foot       & 23 / 7  & Leg chain end-effector \\
\hline
\end{tabular}
\end{table}

\vspace{0.1cm}\noindent\textbf{Analysis.}
We transcribed the audio recordings and segmented the transcripts into smaller units. We then conducted open coding~\cite{boyatzis1998transforming}, applying a constant comparison approach~\cite{corbin2008basics} by systematically comparing each segment with others to identify similarities and group related segments into categories. These categories were further organized into broader themes through iterative analysis. We also employed theoretical sampling~\cite{corbin2008basics} to identify underrepresented or emerging categories and refined them through repeated discussions within the research team.

\begin{table*}[t]
\centering
\caption{\footnotesize Biomechanical metrics with definitions, phase emphasis and grading bands. Upper and lower-body metrics are grouped for compact presentation. All angular values are in degrees; head position in centimetres; width and length values are height-normalised ratios in metres.}
\label{tab:all_metrics}
\resizebox{\textwidth}{!}{%
{\footnotesize
\begin{tabular}{lllccccc}
\hline
\textbf{Metric} & \textbf{Definition} & \textbf{Important Phase} &
\textbf{Elite} &
\textbf{Great} & \textbf{Good} & \textbf{Average} &
\textbf{Poor} \\
\hline
\multicolumn{8}{l}{\textbf{Upper Body Metrics}} \\
\hline
Trunk Inclination ($^\circ$)
& $\angle(\overrightarrow{\text{Pelvis}\rightarrow\text{Spine3}}, \text{Vertical})$
& Pre-Backlift
& 62--77 & 77--85 & 55--62 & 45--55
& $<$45 or $>$85 \\
Shoulder Rotation ($^\circ$)
& Angle between shoulder axis (L/R Shoulder) and pelvis axis
& Backlift to Contact
& 25--55 & 55--70 & 15--25 & 0--15
& $<$0 \\
Front Elbow Angle ($^\circ$)
& $\angle(\text{Shoulder},\text{Elbow},\text{Wrist})$
& All Phases
& 100--130 & 130--145 & 85--100 & 70--85
& $<$70 or $>$145 \\
Head Position (cm)
& Lateral displacement of Head from Pelvis centreline
& All Phases
& $-$10 to 5 & 5--10 & 10--20 & 20--30
& $>$30 \\
\hline
\multicolumn{8}{l}{\textbf{Lower Body Metrics}} \\
\hline
Front Knee Angle ($^\circ$)
& $\angle(\text{Hip},\text{Knee},\text{Ankle})$
& Backlift to Contact
& 140--155 & 155--165 & 130--140 & 120--130
& $<$120 or $>$165 \\
Stance Width (m)
& Euclidean distance between L/R Ankles
& All Phases
& 0.40--0.55 & 0.35--0.40 or 0.55--0.60
& 0.30--0.35 & 0.25--0.30
& $<$0.25 or $>$0.60 \\
Hip Rotation ($^\circ$)
& Angle between hip axis (L/R Hip) and batting axis
& Backlift to Contact
& 30--55 & 55--65 & 20--30 & 10--20
& $<$10 \\
Stride Length (m)
& Anteroposterior distance between leading and trailing ankles
& All Phases
& 0.20--0.45 & 0.45--0.55 & 0.55--0.65 & 0.65--0.80
& $<$0.20 or $>$0.80 \\
Front Foot Orientation ($^\circ$)
& Angle between leading foot direction and batting axis
& All Phases
& $-$45 to $-$35 & $-$35 to $-$25 & $-$25 to $-$15
& $-$15 to $-$5
& $>$$-$5 or $<$$-$45 \\
\hline
\end{tabular}
}}
\end{table*}
\subsection{Findings from Formative Interviews}

\noindent\textbf{Coaching practice remains largely observational, with limited access to quantitative analysis.} Participants relied on observation or slow-motion video replay and verbal feedback rather than quantitative data, consistent with prior work~\cite{cushion2010coach, gilbert2004learning}. As $P9$ noted, \emph{``Right now, we don't have any technique [tool]. We just tell verbally--keep your head straight, shoulder down to the ball.''} These limitations, especially at grassroots and academy levels, highlight the need for low-cost, accessible biomechanical feedback~\cite{worsey2019systematic, barris2008review}.

\vspace{0.1cm}\noindent\textbf{Technique analysis happens in training, not during matches.} All participants noted technique analysis occurs during training rather than competition, consistent with prior literature~\cite{cushion2006understanding}. 
In fact, $P3$ warned that \textit{if you go into the match thinking [about changing a technique], it's over --- you're not [going to] score runs. You [need] 6 months to change [a technical habit].''} This suggested a need to persist analytic artifacts over time for a comprehensive review in the future.

\vspace{0.1cm}\noindent\textbf{Communicating the magnitude of a correction is a challenge.}
Several participants identified the inability to precisely quantify how much a joint needs to move as a core difficulty, echoing challenges in translating biomechanical data into actionable coaching cues~\cite{glazier2010movement, davids2008dynamics}. $P1$ described workarounds involving frame-by-frame manual annotation in analysis software, noting that shot-type categorisation still requires manual tagging. $P3$ observed that the credibility of a correction depends on demonstrating it physically rather than merely describing it. The exact displacements would make it significantly easier to communicate precise correction magnitudes to players. 

\vspace{0.1cm}\noindent\textbf{Coaches distinguish objective biomechanical errors from subjective stylistic preferences.} Participants viewed technique as a mix of objective fundamentals and subjective style. While philosophies varied, they consistently identified body balance, stable head position, and joint alignment as universal fundamentals. $P3$ noted that \textit{You cannot tell an unorthodox player to bat like Virat Kohli. You have to establish a baseline that matches their style.''} This motivated low-level objective kinematic metrics while preserving high-level coach control.

\subsection{Design Goals}
\label{sec:design_goals}

We derived eight design goals from our formative interviews and our own assessment of the capabilities we aim to support in PoseForge.

\vspace{0.10cm}\noindent\textbf{G1. Track identities across multi-person scenes -} 
The system should detect and tract the target athlete in scenes containing other individuals, ensuring the reconstructed motion reflects a single continuous subject.

\vspace{0.10cm}\noindent\textbf{G2. Reconstruct and visualize kinematic motion data -} 
The system should synchronize raw video with reconstructed 3D skeletal motion, supporting interactive spatial exploration of the same.

\vspace{0.10cm}\noindent\textbf{G3. Support metric-grounded technique assessment -} 
The system should compute kinematic metrics graded against baseline ranges, presenting them in a format that is easily interpretable by coaches and athletes without a biomechanics background.

\vspace{0.10cm}\noindent\textbf{G4. Facilitate domain-specific metric editability and extensibility -} 
The system should allow users to customize existing metrics or define new ones that align with their coaching practices and inquiries.

\vspace{0.10cm}\noindent\textbf{G5. Enable a kinematically plausible corrective simulation -} 
The system should allow practitioners to perform ``what-if'' analyses by simulating posture corrections that respect human anatomical constraints.

\vspace{0.10cm}\noindent\textbf{G6. Support multiple interaction modalities -} 
The system should support multiple interaction modalities (e.g., direct manipulation, natural language) to support expert as well as novice users.

\vspace{0.10cm}\noindent\textbf{G7. Close the loop between feedback and quantitative intervention -} 
Coaching suggestions should translate directly into actionable posture corrections, allowing users to apply and refine quantitative interventions seamlessly within a single analysis environment.

\vspace{0.10cm}\noindent\textbf{G8. Maintain transparency, user control, and workflow states -} 
Automated actions should be inspectable, reversible, and overridable. The system must support session persistence, allowing users to save, compare, and manage edited motion sequences over time.

\subsection{Technical Implementation}
To achieve these design goals, we first describe the technical system architecture (Figure~\ref{fig:architecture}), comprising four layers as described next.

\subsubsection{Pose Reconstruction and Identity Tracking}
\label{sec:tech_implementation}
Raw footage is submitted through an upload endpoint that dispatches the video to a multi-person 3D pose estimation pipeline. The backend produces two artefacts per frame: a skeletal joint array in camera-relative 3D coordinates\cite{ferguson2025mhr} and a textured mesh\cite{yang2026sam3dbodyrobust}. Joint coordinates are transformed into a pelvis-centred coordinate system and remapped to leading/trailing-side labels according to batter handedness. This representation is used for metric computation, editing, and AI feedback. A central challenge in multi-person scenes is maintaining stable identity assignment for the batter across frames that may also contain umpires or fielders~\cite{ristani2016performance, bewley2016simple}. Detector-assigned identities are often inconsistent across frames due to occlusions and viewpoint changes~\cite{milan2016mot16}. To address this, we employ a greedy spatial proximity tracker that propagates identity based on temporal coherence in 3D space~\cite{andriluka2018posetrack}.

Detected subjects are matched greedily across frames using pelvis-centroid Euclidean distance with a maximum assignment threshold~(\textbf{G1}). Unmatched detections initialise new tracks, while unmatched tracks are retained or discarded based on temporal continuity. The complete formulation is provided in Appendix~\ref{appendix:tracking}.

Note that PoseForge's key feature is to enable coaches to perform "what-if analysis" on the reconstructed pose and not perform pose reconstruction itself. To perform pose reconstruction, PoseForge utilizes MHR which reports a masked data2model fitting error of ~4.1mm on the 3DBodyTex benchmark. As of November 2025, MHR is the state of the art in parametric human modeling~\cite{ferguson2025mhr,atlas}. As a result, PoseForge's pose reconstruction accuracy is exactly MHR's accuracy.

The frame viewer also renders a 2D pose overlay generated using Google's MoveNet pose estimation model\footnote{\url{https://blog.tensorflow.org/2021/05/next-generation-pose-detection-with-movenet-and-tensorflowjs.html}}, through TensorFlow.js for comparison with the reconstructed 3D skeleton.

\vspace{0.1cm}\noindent\textbf{Skeletal topology and joint hierarchy.}
The MHR pipeline~\cite{ferguson2025mhr} estimates positions for 127 
joints, but the system operates on a reduced set of 19 
anatomically meaningful keypoints organised into a rooted 
kinematic tree~\cite{loper2015smpl}. The root is the Pelvis 
(J0), from which two chains descend: a spinal chain through 
Spine1 $\rightarrow$ Spine2 $\rightarrow$ Spine3 $\rightarrow$ 
Neck $\rightarrow$ Head, and bilateral leg chains through 
Hip $\rightarrow$ Knee $\rightarrow$ Ankle $\rightarrow$ Foot. 
From Spine3, bilateral arm chains branch outward through 
Shoulder $\rightarrow$ Elbow $\rightarrow$ Wrist. 
Table~\ref{tab:keypoints} lists the 19 keypoints, their MHR 
indices, and their role in the constraint pipeline.

Bone lengths $\ell_{ij}$ for each connected pair $(i,j)$ are 
computed from the first frame of the session and cached 
as the rest lengths that subsequent IK solves and 
bone-length preservation passes enforce. This reference 
measurement makes the constraint formulae in 
Appendix ~\ref{appendix:formula} independent of 
stature and avoids re-parameterisation across 
sessions~\cite{aristidou2011fabrik}.

\subsubsection{Kinematic Metric Computation and Grading}

At each frame, the system computes a fixed set of upper- and lower-body metrics from the 3D joint array. Metric definitions are summarised in Table~\ref{tab:all_metrics}. Joint-angle metrics are computed from the angle between adjacent skeletal segments, rotational metrics are computed from the orientation of bilateral joint axes relative to the pelvis-centred coordinate frame, and distance-based metrics are computed using Euclidean distances between the corresponding keypoints. Distance-based measures are normalised by the athlete’s standing height (estimated from the hip-to-head chain) to ensure comparability across players~(\textbf{G3})~\cite{davids2008dynamics, dufek1990biomechanical}. Grading bands (Table~\ref{tab:all_metrics}) are derived from prior 3D kinematic analysis of the off-side front foot drive~\cite{stuelcken2005offside}.

Metrics are evaluated within their most diagnostically relevant movement phases~\cite{ferdinands2013kinematics}: \textit{Pre-Backlift} (setup), \textit{Backlift to Contact} (downswing), and \textit{All Phases} (throughout motion). Each metric is mapped to a five-level ordinal scale using these ranges.
Baseline metric curves from the original motion are rendered as dashed overlays alongside edited curves, enabling visualisation of the applied corrections.

\subsubsection{Custom Metric Definition}
\label{sec:custom_metrics}

To support diverse coaching needs and shot variations~\cite{reid2007coaching}, the system provides a metric editor for defining new metrics or modifying existing ones~(\textbf{G4}). A metric specification includes the type (joint angle or inter-joint distance), selected joints (Table~\ref{tab:keypoints}), editable grading thresholds, an optional phase mask, and a display label. 

Custom metrics inherit the same temporal visualization, phase-aware rendering, and feedback generation as the built-in metrics (Table~\ref{tab:all_metrics}). Existing metrics can also be reconfigured—for example, adjusting Elite thresholds for junior players—without changing the default definitions.

\begin{figure*}[!ht]
    \centering
    \includegraphics[alt={3D reconstructed athlete skeleton across multiple frames. Original and corrected poses are shown as translucent overlays. A cursor is shown dragging a joint's keypoint at a selected keyframe, illustrating how spatial corrections propagate dynamically across the motion sequence.}, width=\linewidth]{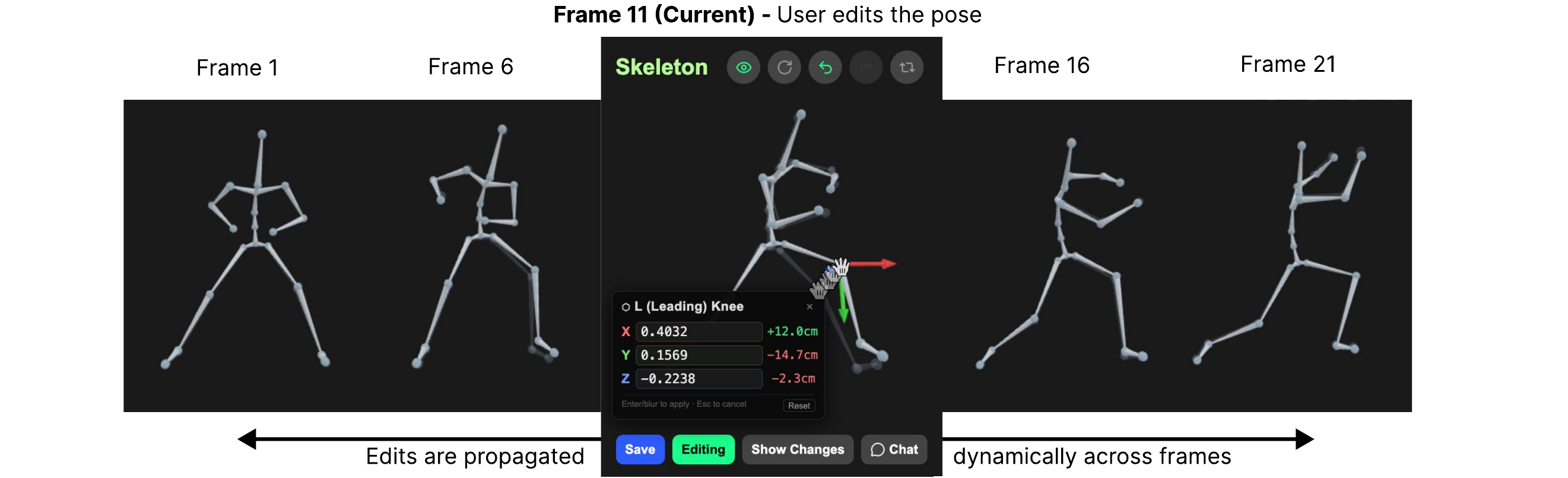}
    \caption{Interactive 3D visualization with joint-level editing controls: corrected pose (silver) is overlaid on the original motion (gray); edits applied at selected keyframes are propagated dynamically across the motion sequence to produce temporally coherent, anatomically plausible corrections.
}
    \label{fig:edit}
\end{figure*}

\subsubsection{Pose Editing with Kinematic Constraints}
\label{sec:skeleton_editing}

To enable corrective simulation, experts can manipulate joint 
positions directly in the skeletal structure through the editing 
interface described in Section~\ref{sec:ui_3d} ~(\textbf{G5}). A displacement applied to a selected keyframe is propagated to all frames within the affected motion interval dynamically before constraint resolution, producing a temporally coherent correction sequence rather than an isolated pose edit. All edits are 
subject to a layered constraint pipeline applied after each drag 
event.

\vspace{0.1cm}\noindent\textbf{Forward And Backward Reaching Inverse Kinematics (FABRIK). }The core of the constraint pipeline is the FABRIK algorithm~\cite{aristidou2011fabrik}. Independent FABRIK chains are maintained for the spine, arms, and legs using bone lengths. When a joint is edited, the corresponding chain is solved iteratively while preserving anatomical structure. The formulation is provided in Appendix~\ref{appendix:fabrik}.

\vspace{0.1cm}\noindent\textbf{Bone-length preservation. }Following inverse kinematics, all connected skeletal segments are re-normalised to their session-specific rest lengths to eliminate numerical drift, as formulated in Appendix~\ref{appendix:bonelength}.

\vspace{0.1cm}\noindent\textbf{Hinge joint limits. }Elbows and knees are constrained to physiologically feasible angular ranges after each inverse-kinematics iteration. Joint-angle computation and clamping are described in Appendix~\ref{appendix:hinge}.

\vspace{0.1cm}\noindent\textbf{One Euro Filter for temporal smoothing. }Edited trajectories are smoothed using the One Euro Filter~\cite{casiez2012oneeuro} to reduce jitter while preserving responsiveness. Parameters were empirically tuned for cricket batting motion. The filter formulation is provided in Appendix~\ref{appendix:oneeuro}.

\vspace{0.1cm}\noindent\textbf{Reference-motion pullback. }To prevent unrealistic deviations from the original motion, edits are progressively blended back toward the reference trajectory as temporal distance from the edited keyframe increases. The formulation is provided in Appendix~\ref{appendix:pullback}.

\vspace{0.1cm}\noindent\textbf{Foot-plant stabilisation. }Detected planted-foot segments are stabilised toward their original positions to preserve ground contact during editing. The formulation is provided in Appendix~\ref{appendix:footplant}.

\vspace{0.1cm}\noindent\textbf{Pipeline composition. }After each drag-release event, PoseForge executes the following stages: (1) FABRIK solving, (2) bone-length preservation, (3) hinge-joint clamping, (4) temporal smoothing, (5) reference-motion regularisation, (6) foot-plant stabilisation, and (7) Laplacian trajectory regularisation. This ordering was selected to preserve anatomical validity while maintaining temporal coherence. Detailed formulations are provided in Appendix~\ref{appendix:pipeline}.

\subsubsection{Natural Language and AI-Assisted Editing}

To support diverse users --- analysts, amateur players, and fans --- the system incorporates a \textbf{feedback agent} that interprets kinematic metrics and generates phase-aware coaching suggestions, and a \textbf{pose editing agent} that converts natural-language instructions into executable joint-displacement specifications~(\textbf{G6}) (Appendix~\ref{appendix:prompt}). Both agents are powered by \texttt{llama-3.3-70b-versatile}~\cite{touvron2023llama}, served via the Groq API~\cite{groq2023} and orchestrated using LangChain~\cite{chase2022langchain}.

\vspace{0.1cm}\noindent\textbf{Feedback Agent.} This agent is triggered when a metric is graded Average or Poor within its phase window (Table~\ref{tab:all_metrics}). For each triggered metric, the system passes its name, value, grade, phase, handedness, and semantic description to the LLM. It uses a structured prompt~\cite{lee2026vistar} encoding metric semantics, phase boundaries, and grade ranges used during metric computation, grounding feedback in the same biomechanical criteria, to generate concise coaching suggestions following augmented feedback principles~\cite{sigrist2013augmented}. Responses are constrained in length and format. Suggestions are generated asynchronously per metric, each rendered inline with an Apply button that forwards the request to the pose editing agent and constraint pipeline (Section~\ref{sec:skeleton_editing}).

\vspace{0.1cm}\noindent\textbf{Pose Editing Agent.}
This agent accepts typed/voice input and converts it into a JSON object. The prompt encodes joint definitions (Table~\ref{tab:keypoints}), current 3D joint states, metric context (Table~\ref{tab:all_metrics}), axis semantics, and safe edit constraints ($0.005$--$0.06$\,m). The model accordingly reasons over the user instruction and current biomechanical state when generating edits. Outputs are validated against kinematic constraints: infeasible edits are clipped and flagged. Both agents maintain a short conversation history to resolve references across successive instructions~\cite{brown2020language}.

\subsubsection{Implementation}
PoseForge is a React-based~\cite{react2013} web application. The 
3D pose visualisation and mesh rendering pipeline is implemented using Three.js~\cite{cabello2010threejs}, with 
GLTF models loaded via the GLTFLoader extension. Natural language 
voice input is handled through the Web Speech 
API~\cite{w3c2012speech}.

\subsection{Design Process and User Interface}
To achieve our design goals, we iteratively explored how to structure, visualize, and interact with the kinematic data. We began with low-fidelity hand-sketches and Figma wireframes to quickly refine layout flows, followed by rapid software prototyping to evaluate library capabilities and technical feasibility. The resulting \textit{PoseForge} user interface (Figure~\ref{fig:teaser}), comprising five main views is described below, followed by alternate design considerations and key trade-offs in Section~\ref{label:alternate-design-considerations}.

\subsubsection{Session Configuration Panel}

The top-left panel handles session setup as shown in Figure~\ref{fig:teaser}(a). An upload control accepts a batting video and, on completion, populates a list of detected persons for selection, ensuring the practitioner analyses a consistent individual rather than an artefact of multi-person detector ordering~(\textbf{G1}). Practitioners then specify batsman handedness, shot type from a vocabulary of 24 cricket strokes, camera view, and athlete anthropometrics before triggering downstream processing.

\subsubsection{Frame Viewer Panel}

The frame viewer presents the decoded video frame at the current playback position alongside playback controls for jump-to-start, jump-to-end, single-frame stepping, and play/pause as shown in Figure~\ref{fig:teaser} (b). A horizontal slider maps the full frame range to a drag target and acts as a cross-panel temporal filter: advancing it or clicking any metric trend curve simultaneously updates the video frame, the 3D joint scene, and the cursor on every metric plot, keeping video evidence and reconstructed motion synchronised~(\textbf{G2}).
A Show Pose Overlay toggle allows users to display or hide the detected 2D keypoints over the video frame, providing an optional visual reference during inspection.
\vspace{1mm}
\begin{figure}[!h]
    \centering
    \includegraphics[alt={Metrics panel visualizing biomechanical measurements and quality grades. Data is plotted on a line chart. The interface includes a 'Show Suggestion' button that reveals an AI-generated coaching tip, and an 'Apply Change' button located to its right that applies the tip to the skeleton.
}, width=\linewidth]{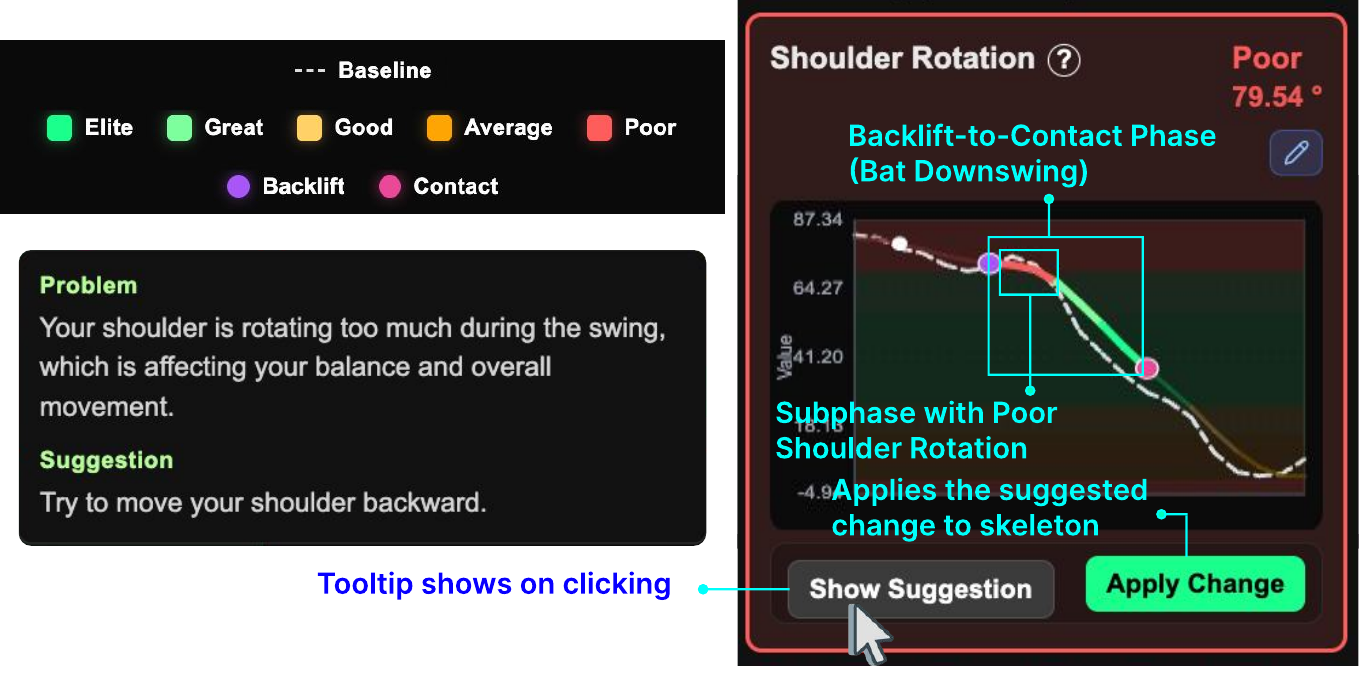}
    \caption{Metric Panel with AI Suggestion: Visualization of computed kinematic metrics with grading and temporal trends, accompanied by context-aware coaching suggestion generated to guide corrective action.}
    \label{fig:inline}
\end{figure}

\subsubsection{3D Visualisation Panel}
\label{sec:ui_3d}

The 3D panel is the primary surface for skeletal inspection and editing. It renders a Three.js scene showing either the reconstructed skeleton---joint spheres connected by tapered bone cylinders---or a textured \texttt{.glb} mesh, toggled via a view-switch control (\faIcon{exchange-alt}) as shown in Figure~\ref{fig:teaser} (c). A ghost overlay (\faIcon{eye}/\faIcon{eye-slash}) renders the original unedited skeleton at reduced opacity alongside the edited version, providing a continuous spatial reference for applied corrections. Arcball orbit, roll, and zoom are shared across both views so that switching preserves the current viewpoint. A reset control (\faIcon{sync-alt}) restores both the viewpoint and joint positions to their original values.

In view mode, hovering over a joint displays a pop-up naming it by its batting-side label and listing the metrics that involve it. In edit mode, clicking a joint renders a three-axis gizmo; dragging an axis translates the joint through the full IK constraint pipeline, ensuring kinematic plausibility~(\textbf{G5}). Edits are propagated dynamically across the motion sequence through inverse kinematics, producing smooth, coherent corrections while preserving the user-specified modification at the edited keyframe, as shown in Figure~\ref{fig:edit}. A numeric keyframe panel in the lower-left corner shows the selected joint's $(x,\,y,\,z)$ coordinates and signed per-axis displacement in centimetres, and accepts direct numeric input for precise corrections. The AI chat assistant (\faIcon{comments}) exposes the same editing capability through natural language and voice input for practitioners who prefer domain-familiar coaching language~(\textbf{G6}).

Undo (\faIcon{undo-alt}) and Redo (\faIcon{redo-alt}) controls step through a joint-array snapshot history. A Show Changes overlay lists, for each modified frame, joints that deviated from the original pose by more than 2\,mm, their displacement magnitudes, and signed per-axis deltas in athlete-relative terms. A Save control (\faIcon{save}) opens a naming dialog that persists the current edited sequence to the Saved Works Panel.
\vspace{1mm}

\begin{figure}[!h]
    \centering
    \includegraphics[alt={Dialog box for defining a custom biomechanical metric. The interface features a clickable human body diagram for selecting joints, toggles for choosing angle or distance measurements, a selector for the movement phase, and interactive sliders to assign ranges for each grade.}, width=\linewidth]{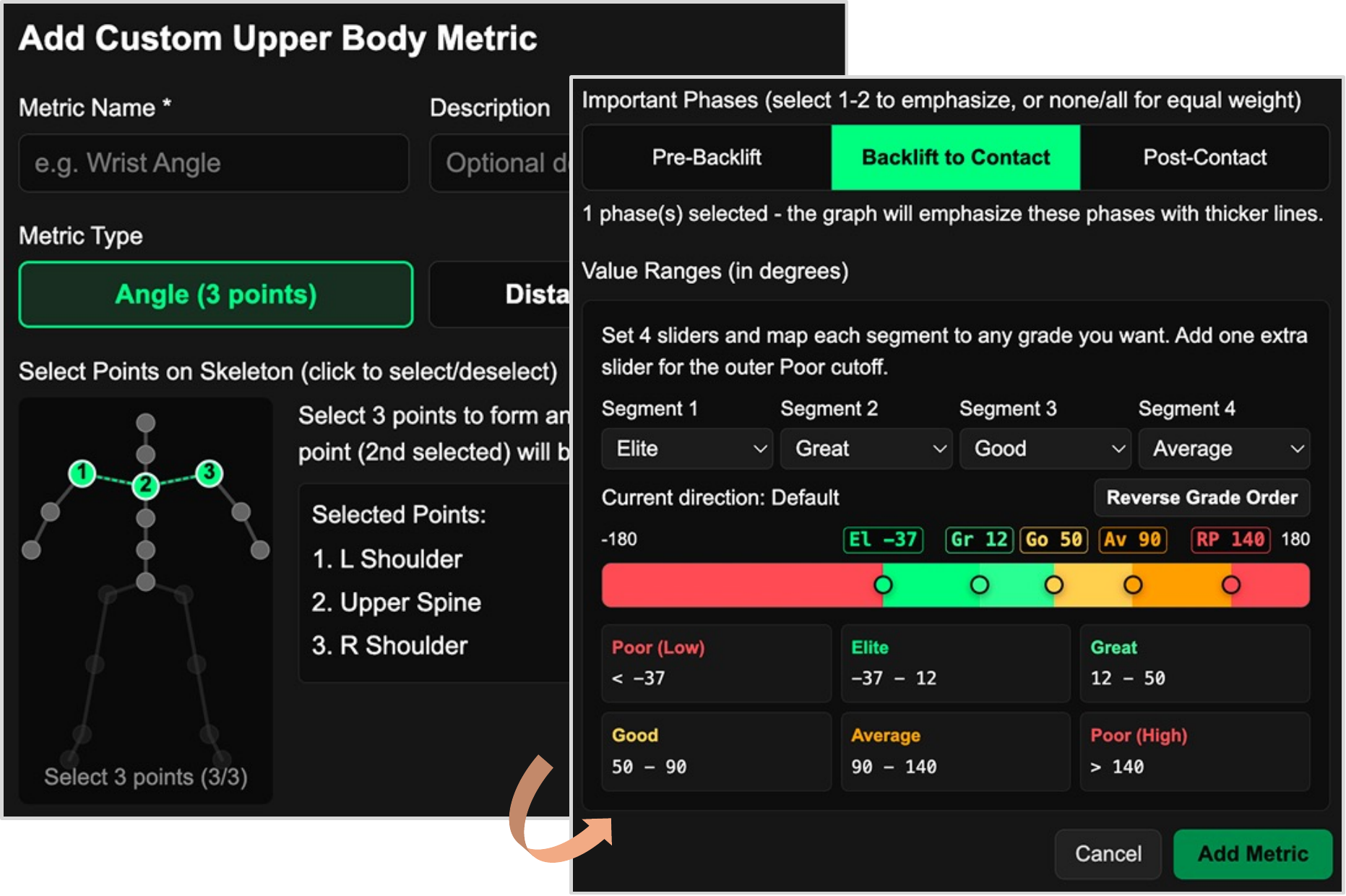}
    \caption{Add Custom Metric Popup: Interface for defining a new kinematic metric by selecting skeletal points, choosing metric type (angle or distance), specifying key movement phases, and configuring grading.}
    \label{fig:custom}
\end{figure}

\subsubsection{Metric Analysis Panel}
\label{sec:ui_metrics}

The metric analysis panel occupies the lower half of the interface, split into Upper Body and Lower Body sub-panels (Figure~\ref{fig:teaser}(d)). Metrics are displayed as cards in a two-column grid, each showing the metric name, current frame value with a five-level grade badge~(\textbf{G3}), a temporal trend curve, and---when graded Average or Poor during its important phase---an inline coaching suggestion with an Apply control that translates feedback into a directly executable joint correction~(\textbf{G7}) (Figure~\ref{fig:inline}). Each trend curve overlays the edited series as a solid colour-graded line against a dashed white baseline~(\textbf{G8}), with grade-band background shading, phase-darkening outside the diagnostic window, and violet/pink markers for the estimated backlift and contact frames respectively. Clicking any curve point synchronises the frame viewer and 3D scene to that frame, while a help control (\faIcon{question-circle}) reveals the metric description and full grade table.

Custom metrics can be added via an Add Custom Metric card (\faIcon{plus-circle}), where practitioners select joints from a schematic skeleton, choose angle or distance type, set grade thresholds on an interactive gradient slider, and optionally restrict evaluation to specific swing phases (Figure~\ref{fig:custom}). Built-in metrics are equally reconfigurable via an edit control (\faIcon{pencil-alt}), enabling norm adjustments for different shot types, age groups, or coaching philosophies~(\textbf{G4}). A legend bar above both sub-panels keys the baseline line, grade colours, and phase markers.

\subsubsection{Saved Works Panel}
The Saved Works Panel provides persistent sequence management as shown in Figure~\ref{fig:teaser} (e). Each saved sequence is a named card showing its title, creation timestamp, and frame count. Clicking a card atomically restores the edited joint array, modification log, and metric state into the active session, immediately updating the 3D scene, trend curves, and coaching suggestions. Sequences are persisted to browser-local storage under a per-athlete key, ensuring that saved work survives page reloads and is scoped to the currently selected person. In addition to the edited pose sequence, PoseForge stores session metadata including the batsman's handedness, shot type (selected from a vocabulary of 24 cricket strokes), camera view, and athlete anthropometrics. On load, the joint array is deep-cloned into the active buffer and the undo history is cleared to prevent cross-sequence state leakage, ensuring that corrected postures can be preserved, revisited, and shown directly to athletes as a reusable coaching reference~(\textbf{G8}).

\subsection{Alternate Design Considerations}
\label{label:alternate-design-considerations}

Throughout our design process, we also navigated several key design trade-offs, as described next.

\vspace{0.1cm}\noindent\textbf{3D Inspection and Editing.} Early prototypes highlighted joints using red and green, but practitioners confused these colors with the system's performance grading. We therefore adopted a neutral grey-and-blue palette for the skeleton, separating editing from performance semantics. We also reduced visual clutter by displaying only the 19 biomechanically relevant keypoints (Table~\ref{tab:keypoints}) instead of all 127 reconstructed joints, improving readability while retaining the information required for coaching and editing.

\vspace{0.1cm}\noindent\textbf{Comparing Original and Edited Motion.} We considered replacing the original motion with the edited sequence after each modification, but practitioners found it difficult to judge the magnitude and direction of corrections from memory. Instead, PoseForge simultaneously visualizes the original and edited poses and overlays the baseline metric trends, enabling direct comparison of posture changes throughout the editing process (\textbf{G8}). A Show/Hide toggle allows practitioners to reveal the original pose when comparing edits or hide it to reduce visual clutter during detailed inspection.

\vspace{0.1cm}\noindent\textbf{Temporal Metric Exploration.} We evaluated both a single chart with toggleable metrics and dense overlapping multi-series plots (\textbf{G2}), but neither supported rapid comparison across metrics. We instead adopted dedicated, always-visible sparkline-style charts for each metric, arranged as a dense grid despite the larger screen footprint. This design preserves an overview of all metrics while allowing practitioners to identify trends and outliers without interaction. To strengthen the connection between metric trends and body movement (\textbf{G3}), every point on a sparkline synchronizes the video frame and 3D pose, allowing practitioners to immediately inspect the corresponding posture.

\vspace{0.1cm}\noindent\textbf{Performance Context in Metric Views.} Since each kinematic metric is only meaningful during specific movement phases, the interface must communicate both timing and performance quality. Tooltips and separate phase tracks either increased interaction or consumed additional space. Instead, we integrated both directly into the charts using background grade bands (Elite--Poor) and dimming outside the metric's diagnostic phase, enabling practitioners to interpret movement timing and performance quality at a glance without additional interaction.

\section{Usage Scenarios}
\subsection{Scenario 1: Self-Directed Learning by a Novice Batter}
Consider Ravi is a 17-year-old club cricketer practising the cover drive independently, without access to a coach. He uploads a side-view clip of him playing cover drive, and browses the metric cards using the tooltip descriptions to understand. The system flags Front Elbow Angle, Front Knee Angle, and Trunk Inclination as Poor or Average.

\vspace{0.1cm}\noindent\textbf{Subscenario 1a: Improving Through AI Suggestions -} Ravi reads the inline suggestion on the Front Elbow Angle card 
--- \textit{``The leading elbow is collapsing too early --- try 
to keep it higher and more extended through the hitting zone''} 
--- and clicks Apply. He works through the remaining flagged 
metrics sequentially, using the ghost overlay after each 
correction to develop a spatial sense of the target posture. 
He saves the corrected sequence, takes it to his next net session 
on a tablet, and re-uploads new footage to track whether his 
live metrics have moved toward the corrected reference, iterating 
across sessions.

\vspace{0.1cm}\noindent\textbf{Subscenario 1b: Side-by-Side Comparison with an 
Elite Player -} After several sessions, Ravi uploads a professional batter's 
cover drive clip, saves it as \textit{``Elite Reference''}, and 
uses the comparison view to render both skeletons simultaneously 
in 3D. At the contact frame, he observes that the elite batter's trunk inclination is more forward, and the shoulder line stays closed longer. He selects Shoulder Rotation 
and confirms that the reference curve stays in the Elite band 
throughout backlift-to-contact while his own breaches Average. 
He adjusts his skeleton toward the reference posture using the 
gizmo controls and saves the result as a corrective target for 
his next training block.

\subsection{Scenario 2: Multi-Metric Comparative Assessment}

Consider Aditya is a sports science analyst preparing pre-tournament 
technical reports on three senior batters. Reviewing the first 
player's straight drive, he finds Stance Width graded Average 
and Front Knee Angle Poor, 
with the angle dropping below $120^\circ$ at contact.

He opens the custom metric editor and defines a new angle 
metric --- \textit{``Back Foot Orientation''} --- using the 
trailing hip, ankle, and foot keypoints, sets the Elite band 
to $25^\circ$--$45^\circ$, and restricts phase emphasis to 
backlift-to-contact. The metric appears in the 
lower-body grid and shows the back foot drifting outside the 
Elite band during the downswing. He repeats the workflow for 
the remaining players, exporting temporal curve snapshots 
from each session and compiling them into a comparative report 
grounded in quantitative grade-band evidence.

\subsection{Scenario 3: Data-driven Season Review}

Consider Sanjay is a batting coach at a professional franchise who uses the system as a longitudinal tracking platform across a six-month season. At the start, he uploads footage of each player's primary strokes under controlled net conditions and saves sessions using the naming convention \textit{``[Player] --- [Shot] --- Pre-Season Baseline''}. Every four weeks, he re-uploads post-match block footage and reloads the baseline session, using the dashed overlay to track metric trajectories over time.

For one player whose front knee angle deteriorates from Good 
pre-season to Poor by the final block, he defines two additional custom metrics - a \textit{``Weight Transfer Index''} and a \textit{``Back Foot Pressure Proxy''} - to investigate the fatigue hypothesis. At season's end, he reviews the full archive of saved sequences in chronological order and exports grade-band annotated curve snapshots for each player, entering pre-season camp with a player-by-player record of which metrics degraded, in which phase, and at what point in the season.

\subsection{Scenario 4: Correcting a Postural Deviation}
Consider Priya, an inexperienced batting coach, observes her player edging cover drives to slips. She uploads footage, selects Shoulder Rotation, and sees a pre-contact peak graded Poor. The feedback suggests: \textit{``The leading shoulder is opening too early --- keep it closed until the ball reaches the hitting zone.''}

She uses the natural-language editor to move the shoulder line back. Using the Show Changes overlay, she verifies adjustments: $+3.2$\,cm anterior-posteriorly and $-1.4$\,cm laterally at the shoulder, with the elbow shifting $+1.8$\,cm anteriorly. Satisfied with the plausibility, she confirms the curve now falls within the Great band.

She saves the sequence as \textit{``Cover Drive --- Shoulder Fix''} and reviews it with the player in the next session. She then reloads it with new footage to check if the correction persists.

\section{Evaluation: Feedback from Cricket Experts}
\label{sec:eval-summative}

To understand how PoseForge can support practitioners in their batting technique analysis workflows, we conducted evaluation sessions with the same eleven cricket experts who participated in the formative study, across the four sessions described in Section~\ref{sec:formative_study}.  Each session included a guided live demonstration of the prototype using footage of a cover drive played by a professional batter, followed by a discussion.

The qualitative data from these discussions were analyzed using the same iterative thematic analysis and constant comparison approach employed in the formative study to ensure consistency across our findings. Participants were asked to reflect on the following topics:

\begin{tcolorbox}[
    breakable,
    colback=gray!20,
    colframe=gray!50,
    boxrule=0.3pt,
    arc=0mm,
    left=1mm,
    right=1mm,
    top=1mm,
    bottom=1mm
]
\footnotesize
\begin{enumerate}[leftmargin=*,nosep]
    \item \textit{``Does PoseForge's joint-based editing workflow match how you reason about correcting a batter's technique?''}
    \item \textit{``How useful and cricket-specific are the AI-generated coaching suggestions? How would you balance them with your own coaching intuition?''}
\item \textit{``Are the provided biomechanical metrics and grading bands sufficient for evaluating batting technique? Would you require additional or customizable metrics?''}
    \item \textit{``Does the edited motion remain realistic after corrections are propagated across frames, and how does this affect your trust in the system?''}
\item \textit{``How well could PoseForge integrate into your existing coaching workflow, and what additional capabilities would improve its practical use?''}
\end{enumerate}
\end{tcolorbox}

\subsection{General Feedback on PoseForge}

\textit{``This system is really good [what you have made], and I think it would really help a lot of coaches at the grassroots level.''} --- $P2$

Overall, participants found PoseForge useful, particularly the quantification of metrics previously assessed only by eye, the ability to do a what-if analysis and communicate corrections through the pose editing interface, and its application in the amateur level. Recurring concerns centred on the dynamics involved in playing the same shot under different conditions and ball types, and the importance of player-specific rather than universal grading norms.

All participants acknowledged the value of moving from 
observational technique assessment to quantified, frame-level analysis. $P1$ stated that while experienced coaches and analysts can pick up small deviations after watching thousands of videos, the tool would be particularly valuable for practitioners who \textit{``don't have an extensive idea of everything --- these tools will definitely help.''} $P2$ also noted that such a system could extend to \textit{``the complete grassroots level, where somebody is just learning to play cricket,''} emphasising its value for players with limited prior technical understanding. $P2$ described the overall experience as a \textit{`game changer,'} noting: \textit{``I've never looked at cricket like this, especially for batting. It has always been listening to coaches and personal experience.''} $P9$ welcomed the tool's potential to improve player outcomes, noting that adoption depends on whether a coach is \textit{``secure enough to want to improve --- if you are, you will use it.''} $P10$ and $P11$ responded positively to the phase-aware temporal curves and the ability to save and reload sessions, identifying the post-session debrief as the primary integration point. 
\subsection{View-Specific Feedback}
\subsubsection{Session Configuration and Frame Viewer Panel}
Participants immediately understood the session configuration parameters as meaningful context for technique assessment. $P2$ raised whether delivery speed and ground conditions should be added as parameters, observing that \textit{``if you see Kohli play the cover drive against someone fast, it would just be a punch --- because there's that much speed on the ball, all he has to do is time it. But in the case of a medium-fast bowler, he has to reach out towards the ball.''} The frame slider and playback controls were universally understood without instruction.

\subsubsection{3D Visualisation Panel}
The skeleton view and ghost overlay elicited the most engaged reactions. $P5$ confirmed the kinematic constraint system produced physically plausible propagated edits: \textit{``You can't just move the knee however you want --- there are constraints applied, and that adds intelligence.''} $P7$ explained that the Show Changes overlay made correction magnitudes explicit, addressing a longstanding communication challenge: \textit{``To be able to say exactly how much the foot has moved --- to that dot, that is actually quite amazing.''} $P3$ observed that demonstrating corrections visually increased player confidence compared with verbal instruction alone: \textit{``When a coach just says it, a player asks, how? But when you show it --- that's when there is belief.''}

Joint editing via the gizmo was described as intuitive by $P3$ and $P5$, though $P7$ noted that non-technical users would not think in joint-by-joint terms: \textit{``I wouldn't go to a player and say increase the angle of your knee. You have to say something they will understand,''} motivating the inclusion of natural language editing alongside direct manipulation. The chat assistant was demonstrated across all sessions using the command \textit{``drop the front wrist by 5 centimetres,''} which $P1$ found to produce realistic whole-body adjustments: \textit{``The front wrist has been dropped, the back elbow has been stretched, and the foot has moved a little bit.``} He consequently preferred natural language interaction over axis-aligned gizmo control for coaching tasks.

\subsubsection{Metric Analysis Panel}
Reactions to the metric cards and temporal curves were positive but accompanied by important nuances. $P1$ raised the most substantive critique, arguing that showing raw angular graphs to a coaching user adds a difficulty: \textit{``Showing shoulder angle is of no value for that user. But how you interpret and display it --- that is going to be important.''} He suggested the system should synthesise multiple metric signals into actionable coaching suggestions in cricket-familiar language: \textit{``When you have 4 or 5 base-level suggestions, your model should be able to give an actionable solution which a batter can actually start to use,''} citing trigger movement timing as the kind of phrase the system should work toward rather than \textit{``increase your elbow angle by 5 degrees.''} This critique was directly addressed by the feedback agent, which translates metric deviations into concise coaching language.

\subsubsection{Custom Metric Editor}
The custom metric editor received uniformly positive reactions across all sessions. $P4$ found the slider interface for defining grading bands flexible and intuitive, suggesting it would allow coaches to encode norms that differ from published reference ranges. $P2$ described the overall editor as \textit{``done amazingly --- you've got the main parts down quite well.''} $P9$ noted that the feature addresses his key reservation about universal metrics: \textit{``These tools will help coaches who have their own set of kinematic evaluation techniques.''}

\subsubsection{Saved Works and Workflow Integration}
All participants identified the Saved Works panel as addressing a gap in their current workflows. $P1$ confirmed that saving and reloading corrected sequences would meaningfully reduce preparation time before player feedback sessions. $P9$, $P10$, and $P11$ identified the post-session debrief as the most natural integration point, with $P10$ noting that reloading a sequence on a tablet at the ground --- rather than emailing a static report --- would represent a meaningful improvement in communication. $P2$ raised the possibility of longitudinal use: \textit{``Moving forward, if you guys can show how a player has improved over time using these sequences, I think that would be very valuable,''} directly validating the persistence and reusability design in G8.

\subsection{Technical Evaluation of AI Agents with Experts}
\label{sec:ai_eval}

Cricket coaching relies on nuanced, often implicit natural-language instructions (e.g., \emph{``don’t let the elbow dip''}), making it essential to evaluate whether such guidance can be accurately translated into precise kinematic edits and actionable feedback. We therefore conducted a technical evaluation with domain experts with two goals: (1) to collect a diverse set of realistic natural-language coaching queries and corresponding pose edits, and (2) to assess whether the system’s outputs align with expert intent. As no objective ground truth exists for pedagogical technique correction for cricket, expert judgment was used as the reference for evaluating correctness.

We invited two cricket experts to interact with PoseForge to perform natural-language-based pose corrections and review generated biomechanical feedback. Each participant interacted with four batting clips (two cover drives and two straight drives) over 20-minute sessions, producing a dataset of \textbf{50 unique pose editing queries} and \textbf{30 unique feedback instances}. This dataset covered a reasonably diverse range of kinematic metrics, deviation magnitudes, batting phases, handedness, and camera viewpoints. Each expert then also compared the system outputs against their intended corrections or ideal feedback and judged whether the agent behavior matched their original intent. We present aggregate findings below, while detailed input queries, system outputs, and expert assessments are in Supplementary Material.

\vspace{0.1cm}\noindent\textbf{Pose Editing Agent.}
For each of the 50 queries, the system-generated JSON output was assessed against the expert-defined ideal correction along five dimensions: joint resolution (JR), axis correctness (AC), magnitude plausibility (MP), frame targeting (FT), and multi-joint coherence (MJC). Each dimension was rated as \textit{Correct}, \textit{Partially Correct}, or \textit{Incorrect}, where \textit{Correct} required agreement within predefined angular ($\pm5^\circ$) or positional ($\pm2$,cm) thresholds where applicable.

Across all prompts, the agent correctly resolved intended joints in \textbf{47/50} cases, achieved correct axis selection in \textbf{45/50} cases, magnitude plausibility in \textbf{37/50} cases, and frame targeting in \textbf{44/50} cases. Multi-joint coherence was evaluated for 14 applicable prompts, with \textbf{9} correct and \textbf{5} partially correct outcomes. Errors primarily stem from ambiguous directional language and challenges in coordinating multi-joint adjustments. Overall, these results suggest that this agent can reliably translate natural-language coaching instructions into plausible kinematic edits while respecting biomechanical constraints.

\vspace{0.1cm}\noindent\textbf{Feedback Agent.}
Each of the 30 system-generated coaching feedback instances was compared against an expert-defined ideal response and assigned an overall compliance rating of \textit{Correct}, \textit{Partially Correct}, or \textit{Incorrect}. Experts additionally rated each response on five-point Likert scales for its \textit{accuracy}, \textit{actionability}, and \textit{clarity}.

Across all cases, the agent achieved \textbf{23/30} correct, \textbf{4/30} partially correct, and \textbf{3/30} incorrect responses. Correct outputs typically identified the underlying biomechanical issue and produced concise, actionable coaching suggestions. Partial and incorrect responses were primarily due to ambiguous directional reasoning or insufficient prescriptive specificity. Overall, experts found the feedback effective in translating biomechanical measurements into structured coaching guidance.

\vspace{0.1cm}\noindent\textbf{Expert Agreement on Ideal Postures.}
To assess consistency in expert-defined targets, participants independently edited and saved their preferred batting poses for each phase. Despite variation in the exact magnitude of corrections (e.g., how far to widen a stance), the experts showed strong consensus on the direction of adjustments (e.g., stance width, head position, knee alignment, shoulder orientation, and balance). Disagreements largely reflected personal coaching styles rather than conflicting biomechanical principles. This supports the notion that while no single canonical "ideal" posture exists in cricket, coaching corrections are firmly grounded in shared biomechanical fundamentals. These findings demonstrate that PoseForge effectively supports objective kinematic analysis while highlighting the need for future systems to accommodate multiple valid stylistic variants.

\section{Discussion}
\noindent\textbf{Democratising Kinematic Analysis using Video.} A consistent reaction across all sessions was that the system's single-video input meaningfully lowers the barrier to entry compared with professional-grade setups that require force plates, inertial measurement units, or motion capture suits. One analyst noted that elite-level analysis currently involves manual keypoint marking in dedicated software --- a time-intensive process requiring specialist knowledge. A head coach observed that local clubs do not have the time or the tools for systematic analysis, and that any tool requiring hardware beyond a smartphone camera would not be adopted at the club level. The system's ability to extract 3D skeletal data and compute kinematic metrics from a single monocular video directly addresses this barrier.

\vspace{0.1cm}\noindent\textbf{Complementing, not Replacing, Coaching Intuition.} A recurring theme was that practitioners do not want a system that replaces their judgement; rather, they want one that provides a concrete foundation for it. One analyst captured this tension directly: metric graphs are valuable as data, but their direct exposure to a coach adds a layer of abstraction that distances the output from the language coaches actually use. PoseForge's feedback agent addresses this challenge by translating metric deviations into concise movement-level coaching phrases. Participant feedback suggests that this translation layer plays a central role in making quantitative biomechanical information accessible and actionable for practitioners without a biomechanics background.

\vspace{0.1cm}\noindent\textbf{{Generalizing Beyond Cricket.}} While PoseForge is a case study on cricket batting, the system's core abstraction--editable human poses and kinematic metrics--has broader applicability. Experts already suggested using PoseForge to other cricket skills such as fast bowling, where more repeatable actions may support more stable kinematic metrics and clearer corrective interventions. The same logic extends to other bat-and-racket sports, and more broadly to movement-centered settings such as rehabilitation or yoga, where practitioners reason about posture, alignment, and controlled motion changes. In each case, the underlying pose-estimation pipeline can provide the reconstructed human motion, while the system designer or domain expert supplies the task-specific metric definitions, threshold ranges, and feedback rules that determine how the movement is graded and interpreted.

\vspace{0.1cm}\noindent\textbf{Takeaways for Visualization and Visual Analytics.} Our findings suggest several design implications for visual analytics systems that support embodied movement analysis. First, quantitative measurements should be translated into domain-familiar terms rather than presented in isolation, so practitioners can immediately relate them to coaching or correction. Second, interactive corrective simulation should be integrated with visual analysis, allowing users to explore hypothetical movement changes rather than only diagnose errors. Third, combining metric-driven reasoning with direct manipulation helps connect quantitative evidence to concrete corrective action. Finally, synchronized views of video, reconstructed motion, metrics, and AI feedback help balance biomechanical detail with cognitive accessibility, enabling practitioners to reason about movement across multiple levels of abstraction. These implications extend beyond cricket to visual analytics for other forms of human-motion analysis.

\section{Limitations and Future Work}

\noindent\textbf{Dual-Pane Synchronized Comparison View.} Participants ($P1$, $P2$) suggested a dual-pane comparison mode for longitudinal self-comparison and elite benchmarking, visualizing metric and motion differences against prior performances or professional references. This could enable LLM agents to propose corrections that progressively bridge the gap to the selected reference.

\vspace{0.1cm}\noindent\textbf{Multi-View Reconstruction.} PoseForge operates on monocular video and therefore inherits limitations of single-view pose reconstruction. $P1$ observed that some biomechanical characteristics like head movement and balance are often easier to evaluate from a front-facing camera angle. Future work could incorporate multi-view capture to improve robustness and reduce ambiguity, while also accounting for contextual factors such as delivery speed, bowling type, and playing conditions.

\vspace{0.1cm}\noindent\textbf{Pelvis-Centred Coordinate Representation.} PoseForge represents motion in a pelvis-centred coordinate system, making biomechanical metrics invariant to the player's global position on the pitch. While appropriate for analysing stroke mechanics, this representation does not preserve absolute translational movement, limiting analysis of batting actions involving substantial movement across the crease, such as stepping out to a spinner or advancing down the pitch. Future work could address this through a hybrid local-global representation that jointly analyses technique and player movement.

\section{Conclusion}
PoseForge is a browser-based system for kinematic analysis and corrective simulation of cricket batting poses.
It reconstructs 3D motion from monocular video, computes phase-aware metrics, and enables interactive editing with AI-driven feedback and correction agents. Evaluations with players and coaches demonstrate its potential to make coaching more quantitative, interpretable, and actionable, while highlighting opportunities for improved transparency and broader metric coverage. 

\section*{Supplemental Materials}
\label{sec:supplemental_materials}
PoseForge is available as open-source software at \url{http://github.com/DataVisards/PoseForge}. The supplemental material includes a video demonstration of PoseForge, the complete evaluation results for the skeleton editing agent (30 prompts), the complete evaluation results for the feedback agent (50 representative metric cases), full prompts and expected outputs, and additional implementation details.

\section*{Figure Credits and Copyrights}
\label{sec:figure_credits}
We have utilized frames from publicly available YouTube 
content published by cricket.com.au (\url{https://www.youtube.com/watch?v=GTPk7_OAhHw}). These materials are used strictly for academic and non-commercial purposes. All rights belong to cricket.com.au.

\acknowledgments{%
    We used Groq to help write and debug portions of \app's source code; AI was not used for data analysis. 
    We thank the members of the DataVisards Lab at HKUST, cricket expert interviewees, and anonymous reviewers for their feedback during various stages of this work.
}

\bibliographystyle{abbrv-doi-hyperref}

\bibliography{template}

\appendix 
\crefalias{section}{appendix} 

\newpage

\section{Mathematical Formulations for Skeletal Estimation and Optimization}
\label{appendix:formula}

This appendix provides the detailed mathematical formulations referenced in Section~\ref{sec:tech_implementation} and Section~\ref{sec:skeleton_editing}. Together, these formulations govern how PoseForge transforms raw 2D videos into temporally stable, anatomically valid 3D kinematic trajectories, and how it handles user-directed corrective edits.

To make the system fully reproducible, our computational pipeline is structured into three sequential optimization stages:

\begin{enumerate}
    \item \textbf{Kinematic Lifting and Optimization:} Translating raw, unconstrained 2D/3D model outputs into a continuous, bone-length-invariant skeletal structure. This stage minimizes projection errors while enforcing hard anatomical constraints.
    \item \textbf{Temporal Smoothing and Dynamics Processing:} Eliminating high-frequency jitter and sensor artifacts inherent to frame-by-frame pose estimation without eroding the explosive acceleration curves typical of athletic movements (e.g., cricket batting strokes).
    \item \textbf{Inverse Kinematics (IK) for Interactive Correction:} Resolving multi-joint configurations during manual or natural language-driven pose editing, ensuring that user-defined joint targets distribute plausibly across the kinematic chain.
\end{enumerate}

The explicit objectives, loss functions, and boundary constraints for each stage are detailed below.
\subsection{Identity Tracking via Greedy Spatial Proximity}
\label{appendix:tracking}

Let $\mathcal{T}_{t-1} = \{ \mathbf{c}_k^{t-1} \}$ denote the set of tracked identity centroids at frame $t-1$, where each centroid $\mathbf{c}_k^{t-1} \in \mathbb{R}^3$ corresponds to the pelvis joint of a tracked subject. Let $\mathcal{D}_t = \{ \mathbf{d}_j^t \}$ denote the set of detected pelvis positions at frame $t$. We compute a pairwise Euclidean distance matrix $D \in \mathbb{R}^{|\mathcal{T}_{t-1}| \times |\mathcal{D}_t|}$ as:

\begin{equation}
\label{eq:tracking_distance}
D_{k,j} = \left\| \mathbf{c}_k^{t-1} - \mathbf{d}_j^t \right\|_2
\end{equation}

\noindent For each existing track $k$, the detection index $j^*$ is selected greedily as:

\begin{equation}
\label{eq:tracking_assign}
j^* = \arg\min_j D_{k,j}
\end{equation}

\noindent The assignment is accepted if the minimum distance satisfies:

\begin{equation}
\label{eq:tracking_threshold}
D_{k,j^*} < \tau
\end{equation}

where $\tau$ is a predefined threshold controlling identity consistency across frames. Each detection is assigned to at most one track, ensuring a one-to-one mapping. Unassigned detections initialise new tracks, while unmatched tracks are either retained or discarded based on temporal continuity. This formulation corresponds to a greedy approximation of bipartite matching~\cite{kuhn1955hungarian} that prioritises spatial proximity over global optimality, and is robust to short-term occlusions and identity swaps in controlled sports analysis settings.

\subsection{Forward and Backward Reaching Inverse Kinematics (FABRIK)}
\label{appendix:fabrik}

The core of the constraint pipeline is the FABRIK algorithm~\cite{aristidou2011fabrik}. Given a kinematic chain of joints $\mathbf{p}_1, \mathbf{p}_2, \ldots, \mathbf{p}_n$ with fixed bone lengths $d_i = \|\mathbf{p}_{i+1} - \mathbf{p}_i\|$ and a target position $\mathbf{t}$ for the end-effector $\mathbf{p}_n$, FABRIK iterates two passes until convergence.

\noindent\textit{Forward pass} --- pull the chain end toward the target, then reposition each joint along the line toward its child at the prescribed distance:
\begin{equation}
\label{eq:fabrik_forward}
\mathbf{p}_n \leftarrow \mathbf{t}, \qquad
\mathbf{p}_i \leftarrow \mathbf{p}_{i+1} +
\frac{d_i}{\|\mathbf{p}_i - \mathbf{p}_{i+1}\|}
(\mathbf{p}_i - \mathbf{p}_{i+1}),
\quad i = n-1, \ldots, 1
\end{equation}

\noindent\textit{Backward pass} --- restore the root to its original position, then propagate forward:

\begin{equation}
\label{eq:fabrik_backward}
\mathbf{p}_1 \leftarrow \mathbf{b}, \qquad
\mathbf{p}_{i+1} \leftarrow \mathbf{p}_i +
\frac{d_i}{\|\mathbf{p}_{i+1} - \mathbf{p}_i\|}
(\mathbf{p}_{i+1} - \mathbf{p}_i),
\quad i = 1, \ldots, n-1
\end{equation}

where $\mathbf{b}$ is the fixed root. The two passes alternate until $\|\mathbf{p}_n - \mathbf{t}\| < \varepsilon$ or a maximum iteration count is reached. The system applies independent FABRIK chains for the spine, each arm, and each leg, with the wrist or foot as the target end-effector. When the total chain length is insufficient to reach $\mathbf{t}$, joints are aligned along the root-to-target direction, preserving bone lengths without distortion~\cite{aristidou2018inverse}.

\subsection{Bone-Length Preservation}
\label{appendix:bonelength}

After each IK solve, a direct constraint pass re-checks every bone defined in Table~\ref{tab:keypoints}. For each connected pair $(i,j)$ with rest length $\ell_{ij}$ cached at session load, the child joint is repositioned along the parent-to-child unit vector:

\begin{equation}
\label{eq:bonelength}
\mathbf{p}_j \leftarrow \mathbf{p}_i +
\ell_{ij} \cdot
\frac{\mathbf{p}_j - \mathbf{p}_i}{\|\mathbf{p}_j
- \mathbf{p}_i\|}
\end{equation}

This ensures that any residual stretch introduced by floating-point accumulation across IK passes is eliminated before the frame is committed.

\subsection{Hinge Joint Limits}
\label{appendix:hinge}

The elbow and knee joints identified in Table~\ref{tab:keypoints} are constrained to physiologically feasible ranges after each IK iteration~\cite{norkin2016measurement}. For a hinge at joint $\mathbf{h}$ with parent $\mathbf{p}$ and child $\mathbf{c}$, interior angle $\theta$ is computed as:

\begin{equation}
\label{eq:hinge_angle}
\theta = \arccos\!\left(
\frac{(\mathbf{p} - \mathbf{h}) \cdot
(\mathbf{c} - \mathbf{h})}
{\|\mathbf{p} - \mathbf{h}\|\,\|\mathbf{c} -
\mathbf{h}\|}
\right)
\end{equation}
and clamped to $[\theta_{\min}, \theta_{\max}]$ (elbows: $8^\circ$--$176^\circ$; knees: $6^\circ$--$178^\circ$). The child is repositioned at the clamped angle within the hinge plane, where the bend axis is recovered from the rest-pose cross product $(\mathbf{p}_{\text{rest}} - \mathbf{h}_{\text{rest}}) \times (\mathbf{c}_{\text{rest}} - \mathbf{h}_{\text{rest}})$ to maintain anatomically consistent joint orientation.

\subsection{One Euro Filter for Temporal Smoothing}
\label{appendix:oneeuro}

Raw joint trajectories across edited frames are smoothed using the One Euro Filter~\cite{casiez2012oneeuro}, a first-order low-pass filter with adaptive cutoff. For a signal value $x_i$ at time step $i$, the filter first estimates velocity using an exponential moving average:

\begin{equation}
\label{eq:oneeuro_velocity}
\hat{\dot{x}}_i =
\alpha_d \cdot (x_i - \hat{x}_{i-1}) \cdot f_s + (1 - \alpha_d) \cdot \hat{\dot{x}}_{i-1}
\end{equation}
where $f_s$ is the frame rate and $\alpha_d$ is the derivative filter coefficient with cutoff $f_c^d$. The adaptive signal cutoff is then:
\begin{equation}
\label{eq:oneeuro_cutoff}
f_c = f_{c,\min} + \beta \cdot |\hat{\dot{x}}_i|
\end{equation}
where $f_{c,\min}$ controls static noise rejection and $\beta$ controls lag under motion. The smoothed signal estimate is:
\begin{equation}
\label{eq:oneeuro_smooth}
\hat{x}_i = \hat{x}_{i-1} + \alpha(f_c) \cdot
(x_i - \hat{x}_{i-1}),
\qquad
\alpha(f_c) = \frac{1}{1 + \tau / T_e},
\quad \tau = \frac{1}{2\pi f_c}
\end{equation}
where $T_e = 1/f_s$ is the sampling period. The system applies independent One Euro instances per joint axis and per metric signal, with parameters $f_{c,\min} \in \{0.5, 1.0\}$\,Hz and $\beta \in \{0.007, 0.01\}$ tuned empirically for batting cadence.

\subsection{Reference-Motion Pullback}
\label{appendix:pullback}

To prevent edits from drifting into kinematically implausible configurations in frames distant from the edited keyframe~\cite{witkin1995motion}, a temporal Gaussian pullback is applied. For each frame $k$ in the affected window centred on the edited frame $k^*$:

\begin{equation}
\label{eq:pullback_weight}
w_k = w_{\min} + (1 - w_{\min})
\cdot \exp\!\left(-\frac{(k - k^*)^2}{2\sigma^2}\right)
\end{equation}
and each non-locked joint position is blended back toward the original rest-pose position $\mathbf{r}_{k,j}$:

\begin{equation}
\label{eq:pullback_blend}
\mathbf{p}_{k,j} \leftarrow
\mathbf{p}_{k,j} +
\lambda_k \cdot (\mathbf{r}_{k,j} - \mathbf{p}_{k,j}),
\qquad
\lambda_k = \lambda_{\text{near}} +
(1 - w_k)(\lambda_{\text{far}} - \lambda_{\text{near}})
\end{equation}
where $\lambda_{\text{near}} = 0.02$ and $\lambda_{\text{far}} = 0.24$.

\subsection{Foot-Plant Stabilisation}
\label{appendix:footplant}

For frames where a foot is estimated to be planted --- identified by comparing ankle and toe velocity against the rest sequence using a threshold of $0.03$\,m/frame, consistent with established foot-contact detection methods~\cite{zeni2008two} --- a position constraint pins the foot joints toward their rest-pose positions:

\begin{equation}
\label{eq:footplant}
\mathbf{p}_{k,j}^{\text{foot}} \leftarrow
\mathbf{p}_{k,j}^{\text{foot}} +
s_k \cdot
(\mathbf{r}_{k,j}^{\text{foot}} -
\mathbf{p}_{k,j}^{\text{foot}})
\end{equation}
where $s_k = s_{\min} + (1 - w_k)(s_{\max} - s_{\min})$ tapers from $s_{\min} = 0.1$ at the edit centre to $s_{\max} = 0.45$ at the temporal boundary. Ankle joints are stabilised at $0.8 \cdot s_k$ to permit natural flexion while keeping the planted foot grounded.

\subsection{Pipeline Composition and Laplacian Regularisation}
\label{appendix:pipeline}

After each drag-release event, the full constraint pipeline executes in order: (1) FABRIK across all kinematic chains in Table~\ref{tab:keypoints} (Eq.~\ref{eq:fabrik_forward}--\ref{eq:fabrik_backward}), (2) bone-length re-enforcement (Eq.~\ref{eq:bonelength}), (3) hinge joint clamping (Eq.~\ref{eq:hinge_angle}), (4) One Euro temporal smoothing (Eq.~\ref{eq:oneeuro_velocity}--\ref{eq:oneeuro_smooth}), (5) reference-motion pullback (Eq.~\ref{eq:pullback_weight}--\ref{eq:pullback_blend}), and (6) foot-plant stabilisation (Eq.~\ref{eq:footplant}), followed by Laplacian regularisation~\cite{sorkine2004laplacian}:

\begin{equation}
\label{eq:laplacian}
\mathbf{p}_{k,j} \leftarrow
\mathbf{p}_{k,j} +
\alpha_{\text{lap}} \cdot
\left(
\frac{\mathbf{p}_{k-1,j} + \mathbf{p}_{k+1,j}}{2}
- \mathbf{p}_{k,j}
\right),
\qquad \alpha_{\text{lap}} = 0.22
\end{equation}

\section{System Prompts for AI Agents}
\label{appendix:prompt}
\begin{promptbox}{Box 1: System Prompt for Feedback Agent}
You are a movement coach generating simple, direct coaching suggestions.

\textbf{Output Format — 2 paragraphs separated by blank line:}

\begin{enumerate}[nosep]
\item \textbf{PARAGRAPH 1 (Problem)}: 1 sentence describing what is wrong using simple observation without jargon.
\item \textbf{PARAGRAPH 2 (Solution)}: 1 sentence with a movement action. Format: [Body part] [direction/movement].
  Use simple verbs: ``move'', ``rotate'', ``lean'', ``extend'', ``bend'', ``keep'', ``hold''.
  Use directions: ``forward'', ``backward'', ``inward'', ``outward'', ``up'', ``down''.
  Prescriptive language preferred: ``Try to...'', ``Focus on...'', ``Make sure to...''
\end{enumerate}

\textbf{Constraints:}
\begin{itemize}[nosep]
    \item Never use technical terms: angles, metrics, elite ranges, technique
    \item Maximum 40 words total
    \item Return exactly 2 paragraphs with 1 blank line between
    \item Respect direction context (left/right based on handedness)
\end{itemize}
\end{promptbox}

\begin{promptbox}{Box 2: System Prompt for Pose Editing Agent}
You are an expert batting biomechanics coach helping users improve their technique by analyzing body movement.

\textbf{Communication Guidelines:}
\begin{itemize}[nosep]
\item Be user-friendly and non-technical—avoid jargon
\item Use clear, observable descriptions instead of technical terms
\item Describe the PROBLEM first, then SOLUTION
\item Use everyday movement language: ``move'', ``hold'', ``rotate'', ``lean'', ``extend'', ``bend''
\item Explain the benefit: balance, vision, power, accuracy, etc.
\end{itemize}

\textbf{Biomechanical Priorities:}
\begin{itemize}[nosep]
    \item Keep edits physically plausible and coordinated
    \item Prefer small, conservative adjustments (0.005--0.06 m)
    \item Maintain balance and stability
    \item Respect shot phase intent: preBacklift, backliftToContact, postContact
\end{itemize}

\textbf{Keypoint Naming Convention:}
\begin{itemize}[nosep]
    \item \textbf{L\_*} = Leading side; \textbf{R\_*} = Trailing side
\end{itemize}

\textbf{Output Format (Strict):} JSON only:\\
\vspace{2pt}
\small
\texttt{\{"explanation": "...", "edits": [\{"keypoint": "...", "axis": "...", "delta": 0.02, "frames": "..."\}]\}}
\end{promptbox}

\vspace{1em}

\section{Technical Evaluation Protocol}
\label{appendix:eval}
This appendix defines the evaluation criteria used for the technical evaluation of the AI agents described in Section~\ref{sec:ai_eval}. Complete prompt-level results for the pose editing agent (50 prompts) and feedback agent (30 test cases) are provided in the supplementary material.

For the \textbf{pose editing agent}, each prompt reports the ideal JSON correction, the agent's generated output, and correctness across five evaluation dimensions:

\begin{itemize}
\item \textbf{Joint Resolution (JR):} Whether the correct joint(s) are identified from the natural-language instruction.
\item \textbf{Axis Correctness (AC):} Whether the direction of motion (x/y/z axis) aligns with the intended correction.
\item \textbf{Magnitude Plausibility (MP):} Whether the magnitude of displacement is biomechanically reasonable.
\item \textbf{Frame Targeting (FT):} Whether the correct temporal segment or phase of motion is selected.
\item \textbf{Multi-Joint Coherence (MJC):} Whether multiple joint edits (if required) are consistent and complete.
\end{itemize}

Each dimension is scored as Correct (\textbf{C}), Partially
Correct (\textbf{P}), or Incorrect (\textbf{I}). A score of
\textbf{C} indicates full alignment with the intended edit,
\textbf{P} indicates minor deviations (e.g., approximate joints
or frames), and \textbf{I} indicates incorrect or missing behaviour.

For the \textbf{feedback agent}, each test case reports the input metric, current value, ideal coaching response, generated response, and an overall compliance score. Compliance is categorized as \textbf{Correct (C)}, \textbf{Partially Correct (P)}, or \textbf{Incorrect (I)}, according to agreement with the expert-defined ideal coaching recommendation.

\label{appendix:feedback}
Each row lists input 
metric and current value, the ideal output, the agent's actual output, and an overall compliance score.
The overall score aggregates as follows:
\textbf{C} = correct, 
\textbf{P} = partially correct, 
\textbf{I} = incorrect.

\clearpage

\end{document}

%% file: packages.tex
\usepackage{xcolor}
\graphicspath{{figs/}{figures/}{pictures/}{images/}{./}} 

\usepackage{tabu}                      
\usepackage{booktabs}                  
\usepackage{lipsum}                    
\usepackage{mwe}                       
\usepackage{longtable}
\usepackage{array}
\usepackage{pdflscape}
\usepackage{mathptmx}                  

\usepackage{enumitem}
\usepackage{soul}
\usepackage{listings}
\usepackage{quoting}

\usepackage{amsmath}
\usepackage{amssymb}   
\usepackage{pifont}    
\usepackage{fontawesome5}
\usepackage[most]{tcolorbox}
\tcbuselibrary{breakable} 
\usepackage{graphicx}
\graphicspath{ {./images/} }
\usepackage{tikz}
\usepackage[percent]{overpic}

\newtcolorbox{promptbox}[1]{
    colback=white,
    colframe=black!70,
    fonttitle=\bfseries,
    title=#1,
    arc=0mm,
    outer arc=0mm,
    left=5pt,
    right=5pt,
    top=5pt,
    bottom=5pt,
    boxrule=0.5pt,
    breakable, 
    enhanced,  
}

\usepackage{float}
\usepackage{xurl}
\tcbuselibrary{breakable}


%% file: commands.tex
\newcommand{\app}{PoseForge\xspace}

\usepackage{setspace} 
\newcommand{\cut}[1]{}

\newif\ifshowfigures
\showfiguresfalse
\newif\ifshowtables
\showtablesfalse